\documentclass[epj]{svjour}
\usepackage{graphics}
\usepackage{MnSymbol}
\usepackage{ulem}
\usepackage{color}
\usepackage{graphicx}
\usepackage{hyperref}   
\usepackage{appendix}
\usepackage{epsfig}
\usepackage{epstopdf}
\usepackage{amsmath}
\usepackage{subfig}
\usepackage{comment}
\usepackage[dvipsnames]{xcolor}
\usepackage{epsfig}
\usepackage{appendix}
\usepackage{graphics}
\usepackage{graphicx}
\usepackage{color}
\newcommand{\ec}{\eta_c}

\newcommand{\old}[1]{}

\newcommand{\be}{\begin{equation}}
\newcommand{\ee}{\end{equation}}
\newcommand{\ba}{\begin{eqnarray}}
\newcommand{\ea}{\end{eqnarray}}
\newcommand{\bi}{\begin{itemize}}
	\newcommand{\ei}{\end{itemize}}

\newcommand{\nn}{\nonumber}

\def\be {\begin{equation}}
\def\ee {\end{equation}}
\def\bea {\begin{eqnarray}}
\def\eea {\end{eqnarray}}
\def\bc {\begin{center}}
	\def\ec {\end{center}}
\def\nn {\nonumber}
\def\eps {\epsilon}
\def\gm {\gamma}

\def\mn {\mu\nu}
\def\al {\alpha}
\def\om{\omega}
\def\({\left(}
\def\){\right)}
\def\[{\left[}
\def\]{\right]}

\def\sumintb{\sum\!\!\!\!\!\!\!\!\!\int\limits}

\begin{document}
\title{The complex heavy quarkonium potential with the Gribov-Zwanziger action}
\author{Manas Debnath\inst{1}
		 \thanks{\emph{email:manas.debnath@niser.ac.in} }
	 \and Ritesh Ghosh\inst{1,2} 
	 \thanks{\emph{email: riteshghosh1994@gmail.com} }
	 \and Najmul Haque\inst{1}
	 		 \thanks{\emph{email:nhaque@niser.ac.in} }
}                     
\offprints{}          
\institute{
	  School of Physical Sciences, National Institute of Science Education and Research,  An OCC of Homi Bhabha National Institute, Jatni, Khurda 752050, India
	  \and 
	   College of Integrative Sciences and Arts, Arizona State University, Mesa, Arizona 85212, USA
  }

\date{Received: date / Revised version: date}
%
\abstract{
	Gribov-Zwanziger prescription in Yang-Mills theory improves the infrared dynamics. In this work, we study the static potential of a heavy quark-antiquark pair with the HTL resummed perturbation method within the Gribov-Zwanziger approach at finite temperature. The real and imaginary parts of the heavy quarkonium complex potential are obtained from the one-loop effective static gluon propagator. The one-loop effective gluon propagator is obtained by calculating the one-loop gluon self-energies containing the quark, gluon, and ghost loop. The gluon and ghost loops are modified in the presence of the Gribov parameter. We also calculate the decay width from the imaginary part of the potential. We also discuss the medium effect of heavy quarkonium potential with the localized action via auxiliary fields.
%
} 
\maketitle
\section{Introduction}
	Quarkonium suppression is one of the signatures of the creation of a novel state of quarks and gluons, i.e., quark-gluon plasma (QGP)~\cite{Muller:1983ed} in relativistic heavy-ion collision experiments at the Large Hadron Collider (LHC) and Relativistic Heavy Ion Collider (RHIC). In 1986, Matsui and Satz~\cite{Matsui:1986dk} suggested that the $J/\psi$ suppression due to Debye screening by color interaction in the medium can play an important role as a QGP formation signature. The heavy quarkonium spectral functions are studied, theoretically, via approaches like Lattice QCD~\cite{Satz:2006kba,Datta:2003ww,Ding:2012sp,Aarts:2012ka,Aarts:2013kaa} and using effective field theories~\cite{Brambilla:2004jw,Bodwin:1994jh,Brambilla:2008cx}. In an EFT, the study of quarkonium potential is not trivial, as the separation of scales is not always obvious. On the other hand, the lattice QCD simulation approach is used where spectral functions are computed from Euclidean meson correlation~\cite{Alberico:2007rg}. As the temporal length at high temperatures decreases, the calculation of spectral functions is not straightforward, and the results highly depend on the discretization effect and have large statistical errors. So, the study of quarkonia at finite temperatures using the potential models~\cite{Mocsy:2008eg,Wong:2004zr} are well accepted and has been investigated widely as a complement to lattice studies.

At high temperatures, the perturbative computations indicate that the potential of the quarkonia state is complex~\cite{Laine:2006ns}. The real part is related to the screening effect of color charges~\cite{Matsui:1986dk}, and the imaginary part is used to obtain the thermal width of the resonance~\cite{Beraudo:2007ky}. So, it was initially thought that the resonances dissociate when the screening is strong enough; in other words, the real part of the potential is too feeble to bind the $Q\bar Q$  pair together. In recent times, the melting of quarkonia is thought to be also because of the broadening of the resonance width occurring either from the gluon-mediated inelastic parton scattering mechanism aka Landau damping~\cite{Laine:2006ns} or from the hard gluo-dissociation process in which a color octet state is produced from a  color singlet state~\cite{Brambilla:2013dpa}. The latter process is more important when the medium
temperature is lesser than the binding energy of the resonance state. As a result, the quarkonium, even at lower temperatures, can be dissociated. 
 Gauge-gravity duality~\cite{Hayata:2012rw,Albacete:2008dz} study also shows that the potential develops an imaginary component beyond a critical separation of the quark-antiquark pair. Lattice studies~\cite{Rothkopf:2011db,Bala:2019cqu} suggest the existence of a finite imaginary part of the potential. Spectral extraction strategy~\cite{Larsen:2019zqv,Bala:2021fkm} and the machine learning method~\cite{Shi:2021qri} have been recently used to calculate the heavy quarkonium (HQ) potential, both indicating a larger imaginary part  of $Q\bar Q$ potential with respect to the pure perturbative QCD calculation. All the studies of heavy quarkonium using potential models have been done within the perturbative resummation framework. In recent times, open quantum systems method~\cite{Brambilla:2020qwo,Akamatsu:2020ypb,Blaizot:2015hya,Brambilla:2017zei,Sharma:2019xum} is drawing more attention to the study of the quarkonium suppression in QGP matter. This method has also been applied to the framework of potential non-relativistic QCD, i.e., pNRQCD~\cite{Brambilla:2008cx,Escobedo:2008sy,Brambilla:2013dpa}.

In the study of hot QCD matter, there are three scales of the system: hard scale, i.e., temperature $(T)$, (chromo) electric $gT$, and (chromo)magnetic scale $(g^2T)$, where $g$ is the QCD coupling. Hard thermal loop (HTL) resummed perturbation theory deals with the hard scale $T$ and the soft scale $gT$ but breaks down at magnetic scale $(g^2T)$, known as Linde problem~\cite{Linde:1980ts,Gross:1980br}. The physics in the magnetic scale is related to the nonperturbative nature. LQCD is appropriate to probe the nonperturbative behavior of QCD. Due to the difficulties in the computation of dynamical quantities with LQCD, it is useful to have an alternate technique to incorporate nonperturbative effects, which can be derived in a more analytical way, as in resummed perturbation theory. A formalism has been developed to incorporate the nonperturbative magnetic screening scale by employing the Gribov-Zwanziger (GZ) action~\cite{Gribov:1977wm,Zwanziger:1989mf} and it regulates the magnetic IR behavior of QCD. 
As discussed in ref.~\cite{Zwanziger:2006sc}, the Gribov parameter is nonzero in the deconfined phase and GZ action relavant in deconfined phase of nuclear matter.
Since the gluon propagator with the GZ action is IR improved, this imitates confinement, causing the calculations to be more consistent with the results of functional methods and LQCD. Equation of state of gluon plasma, thermodynamics and kinetic theories are well studied
within GZ framework at finite temperature~\cite{Zwanziger:2006sc,Begun:2016lgx}. Heavy quark diffusion coefficients~\cite{Madni:2022bea}, quark number susceptibility and dilepton rate~\cite{Bandyopadhyay:2015wua}, quark dispersion relations~\cite{Su:2014rma}, quark-antiquark potential~\cite{Wu:2022nbv}, correlation length of mesons~\cite{Sumit:2023hjj} have been investigated using GZ action. Recent progress can be found in Refs.~\cite{Sobreiro:2005ec,Vandersickel:2012tz,Dudal:2008sp,Capri:2016aqq,Gotsman:2020mpg,Justo:2022vwa,Canfora:2013kma,Dudal:2019ing}.

In this article, we study the heavy quarkonium potential at finite temperatures within a non-perturbative resummation considering Gribov-Zwanziger action, which reflects the confining properties. In this direction, the heavy quark-antiquark potential has been studied at zero temperature in recent times~\cite{Gracey:2009mj,Golterman:2012dx}. We extend the zero temperature calculation to finite temperatures. This is performed from the Fourier transform of the effective gluon propagator in the static limit. In the current article, we have used the Landau gauge and we consider the contribution from the Gribov-modified gluon loop, Faddeev-Popov ghost loop and quark loop in the gluon propagator. Additionally, the contribution from auxiliary fields to the potential is discussed as a possibile source of the linear term in the potential. 
The article is organized as follows: The general structure of the gauge-boson propagator within the Gribov prescription has been constructed in Sec.~\ref{sec:propagator}. The effective gluon propagator is calculated from the one-loop gluon self-energy in Sec.~\ref{sec:oneloop}. In the presence of GZ action, we consider the effect of the Gribov parameter on the gluon and the ghost loop. We compute the real part of potential due to color screening and the imaginary part arising from the Landau damping in Sec.~\ref{sec:hq}. We have also evaluated thermal width for $J/\psi$ and $\Upsilon$. In Sec.~\ref{sec:Effect of Auxilliary fields}, we discuss the medium effect on a local version of the Gribov action where the new set of auxiliary fields is introduced. In the present study, we have incorporated all those effects. Finally, in Sec.~\ref{sec:summary}, we summarize our work.

	\section{General structure of gluon propagator with Gribov parameter}
\label{sec:propagator}
The form of Gribov's gluon propagator in Landau gauge in Euclidean spacetime is given by~\cite{Gribov:1977wm,Zwanziger:1989mf}
\bea
\Delta_{\mu\nu}^0(P)=\left(\delta_{\mu\nu} - \frac{P_\mu P_\nu}{P^2}\right)\frac{P^2}{P^4+\gamma_G^4},
\label{delta00}
\eea
where $\gamma_G$ is the Gribov parameter and $\delta_{\mn}=(1,1,1,1)$. The appearance of the parameter $\gm_G$ in the denominator shifts the pole of the gluon to the unphysical poles i.e. $P^2=\pm i \gm_G^2$. This unphysical excitation using the Gribov parameter indicates the effective confinement of gluons. More detailed discussions about Gribov prescription can be found in review papers~\cite{Vandersickel:2012tz,Sobreiro:2005ec,Dudal:2008sp,Capri:2016aqq}.

The Gribov mass parameter can be obtained from the following gap equation at one-loop order, as~\cite{Vandersickel:2012tz},
\bea
g^2\frac{(d-1)N_c}{d}\sumintb_P \frac{1}{P^4+\gm_G^4}=1,
\eea
where $N_c$ is the number of colors, $d$ is the space-time dimension and $\sumint_P= T\sum_n \int d^3p/(2\pi)^3$. The temporal component of the Euclidean four-momentum possesses discrete bosonic Matsubara frequencies. At asymptotically high temperatures, the form of the Gribov parameter simplifies as ~\cite{Fukushima:2013xsa}, 
\bea
\gm_G&=&\frac{d-1}{d} \frac{N_c}{4\sqrt{2}\pi} g^2 T.
\label{gm}
\eea
In addition to the Gribov parameter, the QCD strong coupling is also needed and we use one-loop running as follows:
\bea
g^2(T)=4\pi \alpha_s=\frac{24\pi^2}{11 N_c-2N_f}\frac{1}{\ln(2\pi T/\Lambda_{\overline{\text{MS}}})},
\eea
where the scale $\Lambda_{\overline{\text{MS}}}=0.176 \rm{GeV}$~\cite{Haque:2014rua}  for $N_f=3$ QCD.


The ghost propagator in the Landau gauge with no-pole condition reads
\bea
D_c(P)=\frac{\delta^{ab}}{P^2} \frac{1}{1-\sigma(P)},
\label{eq:dc}
\eea
where, $(1-\sigma(P))^{-1}=Z_G$ is the ghost dressing function with
\bea
\sigma(P)&=&g^2 N_c \frac{P_\mu P_\nu}{P^2}\sumintb_{K}\frac{1}{K^4+\gm_G^4}\frac{K^2}{(K-P)^2}\nn\\
&\times&\bigg(\delta^{\mn}-\frac{K^\mu K^\nu}{K^2}\bigg).
\eea
Equation~\eqref{eq:dc} in the infrared limit reduces to~\cite{Vandersickel:2012tz}
\bea
D_c(K^2)=\frac{128 \pi^2 \gm_G^2}{g^2N_c}\frac{1}{K^4},
\eea
indicating that ghost propagator is enhanced for $K\rightarrow 0$.


In the vacuum, the gluon self energy can be written as
\bea
\Pi^{\mn}=\Pi(P^2) \Big(\delta^{\mn}-\frac{P^\mu P^\nu}{P^2}\Big)=\Pi(P^2) V^{\mn},
\eea
where $P^\mu=(p_0,\bf{p})$ is the gluon four momentum and $\Pi(P^2)$ is the Lorentz invariant quantity depending upon $P^2$.

In a thermal medium, the general structure of gluon self-energy, satisfying tranversality condition $p_\mu \Pi^{\mn}=0$, can be written in terms of two independent symmetric tensors, as
\bea
\Pi^{\mn}= \Pi_T A^{\mn}+\Pi_L B^{\mn},
\eea
where the forms of tensors are $B^{\mn}= \frac{\bar u^\mu \bar u^\nu}{\bar u^2}$ with $\bar u^\mu=V^{\mn}u_\nu$ and $A^{\mn}=V^{\mn}- B^{\mn}$.  We choose the rest frame of the heat bath {\it{i.e.,}} $u^\mu=(1,0,0,0)$. The form factors $\Pi_L$ and $\Pi_T$ are functions of two Lorentz scalars $p_0$ and $p=|\bf{p}|$. 

The effective gluon propagator can be written using Dyson-Schwinger equation as  
\bea
\Delta_{\mu\nu}^{-1}&=&\left(\Delta_{\mu\nu}^0\right)^{-1}+\Pi_{\mu\nu}.
\label{DS}
\eea
Here we would use the form of the gluon propagator, 
\bea
\Delta_{\mu\nu}^0(P)=\left(\delta_{\mu\nu} - \frac{P_\mu P_\nu}{P^2}\right)\frac{P^2}{P^4+\gamma_G^4}+\xi \frac{P_\mu P_\nu}{P^4},
\label{delta00xi}
\eea
where $\xi $ is the gauge  parameter. This form is useful to invert the propagator. In the final expression of the effective gluon propagator, we put $\xi=0$ (Landau gauge).
The form of the inverse gluon propagator from Eq.~\eqref{delta00xi} reads as
\bea
\left(\Delta_{\mu\nu}^0\right)^{-1}=\left(\delta_{\mu\nu} - \frac{P_\mu P_\nu}{P^2}\right)\frac{P^4+\gamma_G^4}{P^2}+\frac{1}{\xi}P_\mu P_\nu.
\label{Delinv}
\eea
Putting the form of $\left(\Delta_{\mu\nu}^0\right)^{-1}$ in eq.~\eqref{DS}, one gets the expression of inverse effective gluon propagator
\bea
\Delta_{\mu\nu}^{-1}
&=&\frac{1}{\xi}P_\mu P_\nu+\left(\frac{P^4+\gamma_G^4}{P^2}+\Pi_T\right)A_{\mu\nu}\nn\\
&+&\left(\frac{P^4+\gamma_G^4}{P^2}+\Pi_L\right)B_{\mu\nu}.
\eea
To find the effective gluon propagator, one can write the general structure of gluon propagator in the tensor basis as, $\Delta_{\mu\nu} = \alpha P_\mu P_\nu + \beta A_{\mu\nu} + \gamma B_{\mu\nu},$
where we need to find the coefficients $\al, \beta,\gm$. Then we use the relation $\delta^\nu_\alpha=\Delta^{\mu\nu} \left(\Delta_{\mu\alpha}\right)^{-1}$ to obtaine the coefficients $\al, \beta,\gm$ as
\bea
\beta&=&\frac{P^2}{P^4+\gamma_G^4+P^2\Pi_T},\nn\\
\gamma&=&\frac{P^2}{P^4+\gamma_G^4+P^2\Pi_L},\ \ 
\alpha=\frac{\xi}{P^4}.
\eea
So, in-medium gluon propagator with Gribov term can be written as
\bea
\Delta_{\mu\nu} &=& \frac{\xi P_\mu P_\nu}{P^4} +\frac{P^2 A_{\mu\nu}}{P^4+\gamma_G^4+P^2\Pi_T} \nn\\
&+& \frac{P^2B_{\mu\nu}}{P^4+\gamma_G^4+P^2\Pi_L}\nonumber\\
&=&\frac{\xi P_\mu P_\nu}{P^4} + D_T A_{\mn}+D_L B_{\mn}.
\label{gluonprop}
\eea
One can recover the usual Euclidean Matsubara gluon propagator (without Gribov term) by putting $\gamma_G=0$, such that
\bea
\Delta^T_{\mu\nu} = \frac{\xi P_\mu P_\nu}{P^4} +\frac{A_{\mu\nu}}{P^2+\Pi_T} + \frac{B_{\mu\nu}}{P^2+\Pi_L}.
\eea
From the pole of the propagator dispersion relation of gluons can be obtained. In the next section we would find $\Pi_L$ and $\Pi_T$ from one loop gluon self-energy.

\section{One-loop gluon self-energy}
\label{sec:oneloop}
\begin{figure}[tbh]
	\begin{center}
		\includegraphics[scale=.5]{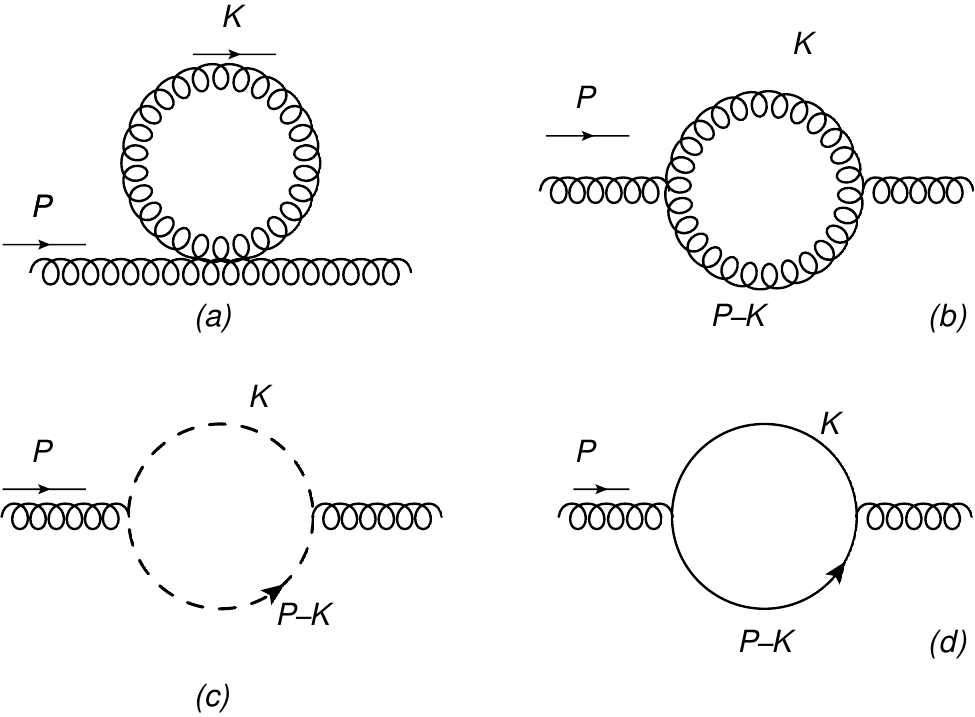}
		\caption{Feynman diagrams for gluon self-energy.  }
		\label{fig:Feynman}
	\end{center}
\end{figure}
In this section, we discuss the one-loop gluon self-energy in Quark-Gluon Plasma in the Gribov quantization scheme at finite temperatures. We can obtain Debye screening mass from the static limit of the polarization tensor. There are four contributions to the gluon self-energy: the tadpole diagram ($\Pi^{(a)}$), gluon loop ($\Pi^{(b)}$), quark loop ($\Pi^{(d)}$), and ghost loop ($\Pi^{(c)}$) as shown in fig.~\ref{fig:Feynman}.
We note that the quark loop is unaffected by the Gribov parameter. 

First, we write down the expression of gluon self-energy from the tadpole diagram
\bea
\Pi_{\mu \nu}^{(a)}(P) =3\, C_A g^2 \delta_{\mu \nu} \int \frac{d^4 K}{(2 \pi)^4} \Delta (K)  = 3 C_A g^2 \delta_{\mu \nu} J_1,
\label{eq:pia}
\eea
with $\Delta(K)=\frac{K^2}{K^4 + \gamma_G^4}.$
$ \Pi_{\mu \nu}^{(a)}$ is independent of external momentum $(P)$  and $C_A=3$. The expression of $J_1$ are derived in appendix~\ref{app_selfenergy}.  The gluon loop contribution in the Landau gauge reads as
\bea
\Pi_{\mu \nu}^{(b)}
= - 6 g^2 C_A I_{\mu \nu} -\frac{1}{2} g^2 C_A \delta_{\mu \nu} J_1 +\frac{1}{2}g^2 C_A \delta_{\mu \nu} J_2,
\label{eq:pib}
\eea
The total contribution from the tadpole and gluon loop can be obtained by adding Eq.~\eqref{eq:pia} and Eq.~\eqref{eq:pib} as
\bea
&&\hspace{-1cm}\Pi_{00}^{a+b}(p_0,\mathbf{p})\nn\\
&=& g^2  C_A \bigg(-\frac{7}{2} J_1 + \frac{13}{2} J_2(p_0,\mathbf{p})+ 6 J_3 (p_0,\mathbf{p})\bigg),
\eea
where the detailed expressions of $J$'s and $I_{\mu \nu}, $ are evaluated in appendix~\ref{app_selfenergy}.

The contribution of the ghost loop to the gluon self-energy can be computed as
\bea 
\Pi_{\mu \nu}^{(c)}(P)&=& C_A g^2 \int \frac{d^4K}{(2\pi)^4} K_{\mu} K_{\nu}D_c (K) D_c (P-K) \nonumber \\
&=& C_A g^2 \left(\frac{128 \gamma_G^2 \pi^2}{N_c g^2}\right)^2 I_{00}^G(p_0,\mathbf{p})
\label{pic}
\eea
The form of $I_{00}^G$ can be found in appendix~\ref{app_selfenergy}.

Quarks are not affected by the Gribov condition as the path integral over the gauge fields are constrained within the Gribov region. So, the quark-loop contribution to gluon self-energy is the same as of the usual HTL case. In our present case, we only need the temporal component of self-energy. So for $N_f$ number of flavors, we have
\bea
\Pi_{00}^{(d)} (p_0,\mathbf{p})=\frac{g^2 T^2 N_f}{12}\left(1-\frac{p_0}{2 p}\log\left[\frac{p_0 + p}{p_0 - p}\right]\right)
\eea
and in the static limit
\bea
\Pi_{00}^{(d)} (p_0\to 0,\mathbf{p})= \frac{g^2 T^2}{6}\left(1- \frac{p_0}{2p }i \pi\right).
\eea
Finally, we need to add up all the contributions to get the total self-energy, i.e., $\Pi_{\mn}^{(a+b+c+d)}(P)$.

\begin{figure}[tbh]
	\begin{center}
		\includegraphics[scale=.54]{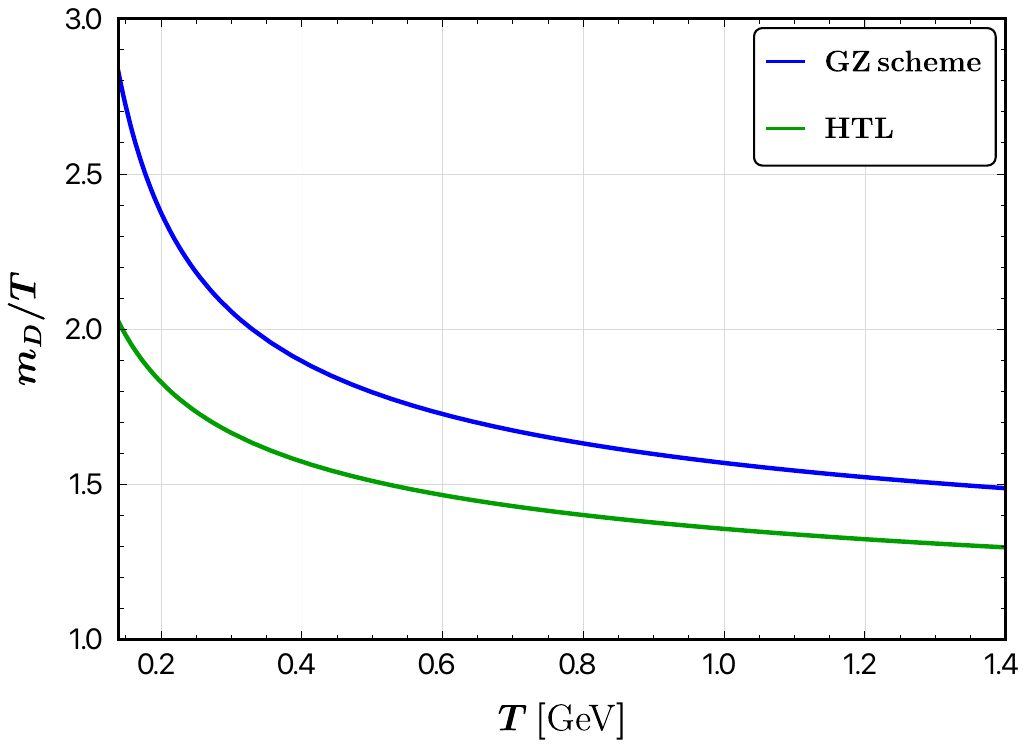}
		\caption{Debye mass scaled with temperature is plotted with temperature.}
		\label{fig_debye}
	\end{center}
\end{figure}

In a thermal medium, the Debye screening mass ($m_D$) is an important quantity as this determines how the heavy quarkonium potential gets screened with inter-quark distance. Debye mass $m_D$ is defined by taking the static limit of $`$$00$' component of the gluon self-energy as $m_D^2=\Pi^{00}(p_0=0,{\bf{p}}\rightarrow 0).$
Within the one-loop order, the Debye screening mass squared can be computed in thermal medium as~\cite{Bellac:2011kqa}
\bea
(m_D^{T})^2=\Big( N_c+\frac{N_f}{2} \Big)\frac{g^2 T^2}{3}.
\eea
The gluon and ghost loop of the gluon self-energy gets modified in the presence of the Gribov medium. By taking the static limit of $\Pi_{00}^{(a+b+c+d)}(P)$, we get the Debye mass in thermal Gribov plasma. We have compared the Debye masses for perturbative and Gribov cases in fig.~\ref{fig_debye}. 
\section{Heavy quark-antiquark potential}
\label{sec:hq}
The physics of quarkonium state at zero temperature can be explained in terms of non-relativistic potential models, where the masses of heavy quark ($ m_Q $) are much higher than QCD scale ($ \Lambda_{QCD} $), and velocity of the quarks in the bound state is small, $v\ll 1$~\cite{Lucha:1991vn,Brambilla:2004jw}. Therefore, to realise the binding effects in quarkonia, one generally uses the non-relativistic potential models~\cite{Eichten:1979ms}. 
In the following subsections, we investigate the effect of the Gribov parameter at finite temperature in the heavy quark-antiquark potential. 
\subsection{Real part of potential}
\label{subsec:real}
\begin{figure}[tbh]
	\begin{center}
		\includegraphics[scale=.5]{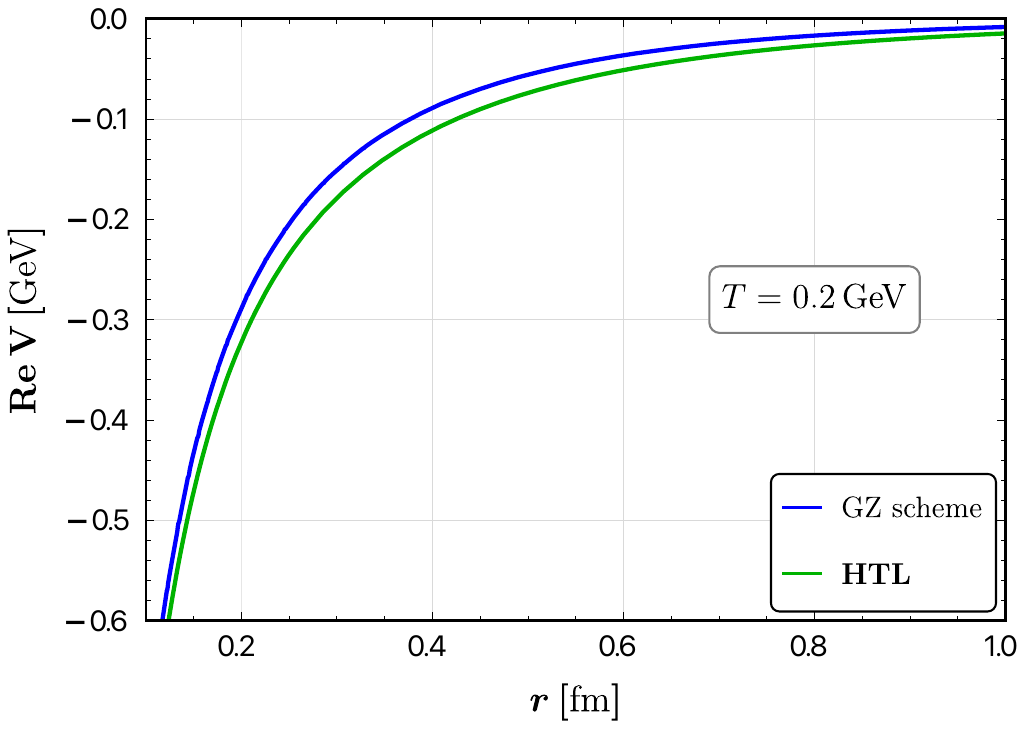}
		\caption{Variation of real part of quarkonium potential with distance for fixed temperature $T=0.2$ GeV. }
		\label{fig_real}
	\end{center}
\end{figure}
The real part of the HQ potential gives rise to the Debye screening, and in the non-relativistic limit, it can be determined from the Fourier transform of static gluon Matsubara propagator as
\bea
V(r)=g^2 C_F\int \frac{d^3p}{(2\pi)^3} \,\,\Delta_{00}(p_0=0, {p})\,\, e^{i \bf{p}\cdot \mathbf{r}}
\eea
For the free case we can write, $\Delta^0_{00}(p_0=0, {p})= -  p^2 /(p^4 + \gamma_G^4 )$
and after the Fourier transform leads to the expression
\bea
V_0(r)= -\frac{C_F g^2}{4 \pi r} \exp\left[- \gamma_G r/\sqrt{2}\right] \cos\left(\gamma_Gr/\sqrt{2}\right).
\eea
To investigate the static potential at finite temperature, we can consider a medium effect to this potential due to thermal bath. In a thermal medium, the effective gluon propagator is given by,
\bea
\Delta_{00}(p_0=0,{p})=  - \frac{{p}^2}{{p}^4 +\gamma_G^4+{p}^2 \Pi_L(p_0=0,{p})} .
\eea
In this case, the heavy quarkonium potential in coordinate space can be written as
\bea
V(r) &=&-g^2 C_F\int \frac{d^3p}{(2\pi)^3} \,\,e^{i \bf{p}\cdot \mathbf{r}}\,\,\frac{p^2}{p^4 +\gamma_G^4+p^2 m_D^2}\nn\\
&\!\!\!=&\!\!\!-\frac{g^2 C_F}{(2 \pi)^3}\int  \, \frac{4 \pi  p^3 \,dp\,}{p^4 +\gamma_G^4+p^2 m_D^2}  \frac{\sin(p\,r)}{r}\nn\\
&=&-\frac{g^2 C_F}{8 \pi\,r}\frac{1}{\sqrt{m_D^4-4\gamma_G^4}}\nn\\
&\times&\bigg[\left(-m_D^2+\sqrt{m_D^4-4\gamma_G^4}\right)e^{-\frac{r}{\sqrt{2}}\sqrt{m_D^2-\sqrt{m_D^4-4\gamma_G^4}}}\nn\\
&+&\bigg(m_D^2+\sqrt{m_D^4-4\gamma_G^4}\bigg)e^{-\frac{r}{\sqrt{2}}\sqrt{m_D^2+\sqrt{m_D^4-4\gamma_G^4}}}\bigg],\qquad
\label{eq:vg}
\eea
where $m_D$ is the Debye mass. The last line simplification is done by contour integral where $m_D\ge\sqrt{2}\gamma_G$.  When the Gribov parameter $\gm_G=0$, from Eq.~\eqref{eq:vg} we get back the isotropic potential
\bea
V(r,\gm_G=0)\!&=&\!-g^2 C_F\int \frac{d^3p}{(2\pi)^3} \,\,e^{i \bf{p}\cdot \mathbf{r}}\,\,\frac{1}{p^2 +(m^T_D)^2}\nn\\
&=&-\frac{g^2 C_F}{4\pi r} \exp[-{r}/{m^T_D}],
\eea
where $m^T_D$ is the Debye mass for HTL case without Gribov.

For the $r\rightarrow 0$ limit, the denominator will be dominated by $p^2$ as $p\rightarrow \infty$ and we recover vacuum Coulomb potential
\bea
V(r\rightarrow 0)&=&-g^2 C_F\int \frac{d^3p}{(2\pi)^3} \,\,e^{i \bf{p}\cdot \mathbf{r}}\,\,\frac{1}{p^2}\nn\\
&=&-\frac{g^2 C_F}{4\pi r}.
\eea

The real part of the potential is shown in fig.~\ref{fig_real}. We compare the Gribov-modified potential with the usual HTL case. The blue line shows the Gribov result using Gribov parameter of Eq.~\eqref{gm} and green line shows the quarkonium potential obtained from the usual HTL calculation.
In the presence of the Gribov parameter, the potential value is less negative than the HTL case at fixed $r$, indicating more screening. This happens because the Debye mass of Gribov plasma is higher than the perturbative medium at a fixed temperature. 

\subsection{Imaginary part of potential}
\label{subsec:im}
\begin{figure}[tbh]
	\begin{center}
		\includegraphics[scale=.5]{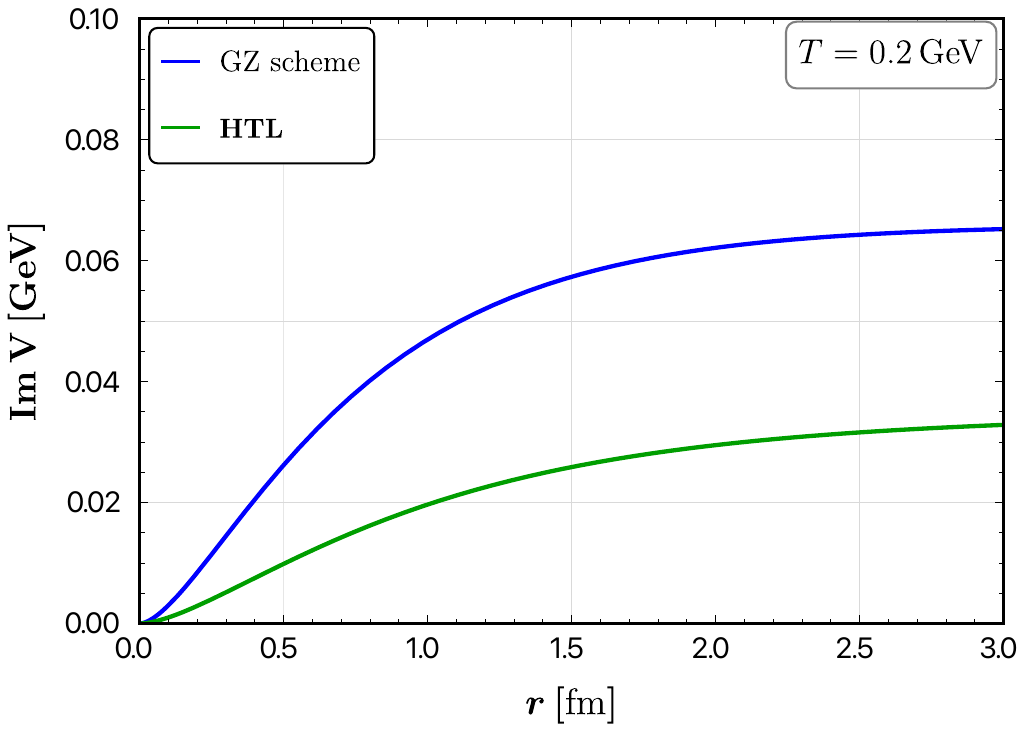}
		\caption{The imaginary part of the quarkonium potential as a function of distance for temperature $T=0.2$ GeV.}
		\label{fig_im}
	\end{center}
\end{figure}
The heavy quarkonium is dissociated at finite temperature by color screening and Landau damping. The Landau damping part is related to the imaginary part of the gluon self-energy as initially suggested in Ref.~\cite{Laine:2006ns}. 
The imaginary part of the in-medium heavy quark-antiquark potential related to the gluon propagator is written as,
\bea
\text{Im} \,V(r) = \int \frac{d^3 p}{(2 \pi)^3} (e^{i \mathbf{p}\cdot \mathbf{r}}-1) \left[ \lim_{p_0 \to 0} \text{Im} \, \Delta_{00}(P) \right].\ \ \ 
\eea

Here one should note that, the Matsubara propagator can be expressed in terms of spectral function as,
\be
\Delta^{\mu \nu}(p_0,p)=\int_{-\infty}^{\infty} \frac{dp_0'}{2\pi} \frac{\rho_i(p_0^\prime,p)P_i^{\mu \nu}}{p_0^\prime-p_0}
\ee
 From which one can express the spectral function as,
\begin{eqnarray}
	\text{Im}\, \Delta^{\mu \nu}(P)=\rho_i P_i^{\mu\nu},
	\label{rho}
\end{eqnarray}
where for Gribov Zwanziger scheme,
\begin{eqnarray}
	\rho_i= P^4\frac{\text{Im}\,\Pi_i}{(P^4+\gm_G^4+P^2 \,\text{Re}\,\Pi_i)^2+P^4 (\text{Im}\,\Pi_i)^2}.
\end{eqnarray}

Hence, the imaginary part of the effective gauge boson propagator can be written as,
\begin{equation}
	\text{Im}\, \Delta^{\mu \nu}(P)= \frac{P^4\,\,\text{Im}\,\Pi_i}{(P^4+\gm_G^4+P^2 \,\text{Re}\,\Pi_i)^2+P^4 (\text{Im}\,\Pi_i)^2} P_i^{\mu\nu},
	\label{rho}
\end{equation}
for $i=T, L$.  From Eq.~\eqref{gluonprop}, $P_{\mu \nu}^i$ are the projection operators.
%

The spectral representation of a quark propagator in GZ scheme within HTL resummation was discussed in \cite{Bandyopadhyay:2015wua,Su:2014rma}. It was shown in \cite{Su:2014rma} that if the Gribov gluon propagator is used to study quark collective behavior, a new space-like mode appears in addition to the two existing HTL modes and sprectral representation is well defined with this new dispersion mode.
We believe one can show similar spectral representation for gluon propagator as well.

The imaginary parts are coming from the Landau damping. The contributions appearing from the self-energy diagrams are finally calculated numerically. In the following subsection, calculations of decay width related to the imaginary part of the potential are presented.
The variation of the imaginary part of the potential with separation distance $r$ is shown in fig.~\ref{fig_im}. The Gribov part is shown with a blue line, whereas the green line denotes the perturbative part. It can be found from the graph that the magnitude of the imaginary part of potential in Gribov is more significant than in the  perturbative case at fixed temperature $T$. This behavior indicates more contribution to the Landau damping-induced thermal width in Gribov plasma obtained from the imaginary part of the potential.

\subsection{Decay Width}
\label{sec:decay}
In the previous subsection, we obtained the imaginary part of the HQ potential by calculating the imaginary part of gluon self-energy. The imaginary part arises due to the interaction occurs between the heavy quarks and particles of momenta $\sim T$, with the exchange of soft space-like gluons, namely, Landau damping. The physics of the finite width emerges from this Landau damping. The imaginary part of the HQ potential plays a crucial role in the dissociation of the HQ-bound state.  The formula which gives a good approximation of the decay width ($\Gamma$) of $Q \bar Q $ states is written as~\cite{Thakur:2016cki},
\begin{equation}
\Gamma(T)= -\int \,|\psi(\mathbf{r})|^2 \, \text{Im}\, V(\mathbf{r, }T) \, d^3 \mathbf{r} .
\end{equation}
Here we have used the Coulombic wave function for the ground state of hydrogen like atom which is given by $\psi(r)= \frac{1}{\sqrt{\pi b_0^3}}\,e^{-r/b_0},$
%
where $b_0= \frac{1}{m_Q \alpha_s}$ is the Bohr radius of the heavy quarkonium system and $m_Q$ is the mass of the heavy quark and antiquark. We determine the decay width by substituting the imaginary part of the potential for a given temperature. We evaluate the decay width of the following quarkonia states, $J/\psi$ (the ground state of charmonium, $c\bar c$ ) and $\Upsilon$ (bottomonium, $b\bar b$).

\begin{figure*}[h]
	\begin{center}
		\includegraphics[scale=.5]{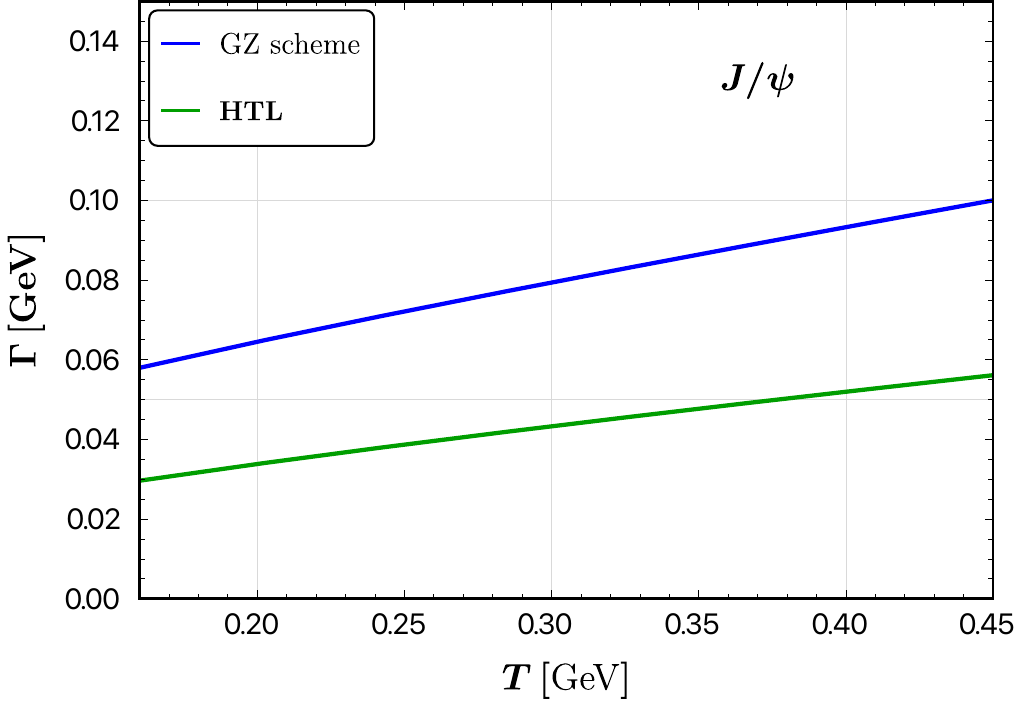}
		\includegraphics[scale=.5]{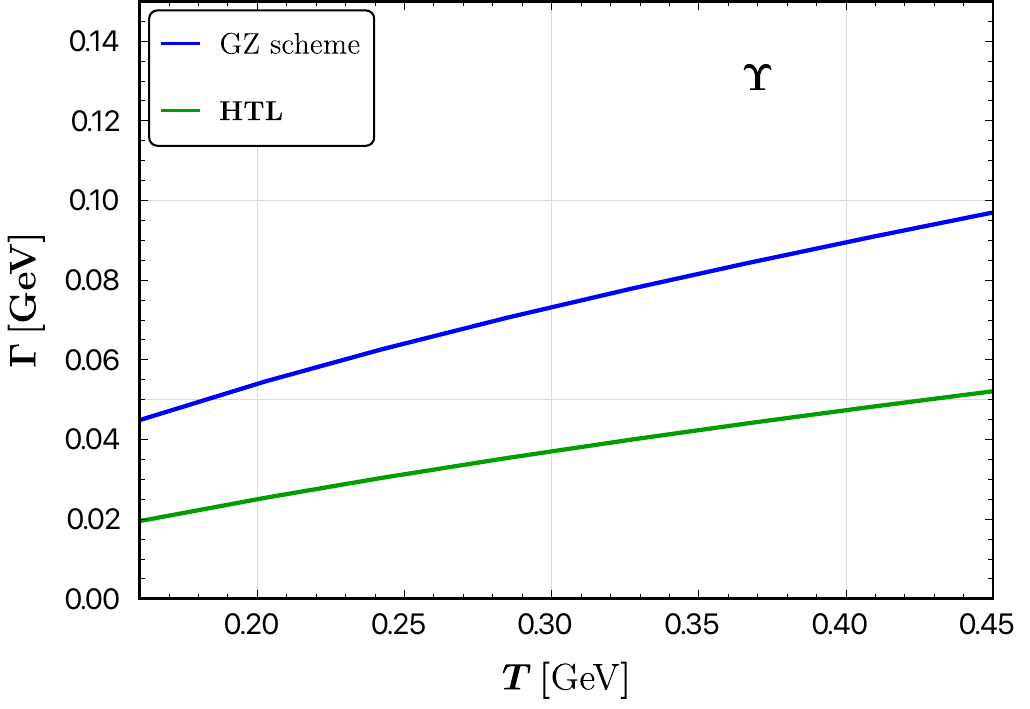}
		\caption{The temperature variation of decay for a charmonium (left panel) and bottomonium (right panel). }
		\label{fig_decay}
	\end{center}   
\end{figure*}

Fig.~\ref{fig_decay} displays the variation of decay width with temperature $T$. In this calculation, we have taken the masses of charmonium $(m_c)$ and bottomonium $(m_b)$ as $m_c=1.275$ GeV and $m_b=4.66$ GeV, respectively. The decay width for the charm (left panel) and the bottom (right panel) are shown. As the magnitude of the imaginary part of the potential in Gribov is larger than usual HTL case, we are getting larger decay width for the case of Gribov plasma. For both the bottom and charm, the decay width is increased in the case of Gribov plasma. The thermal width for $\Upsilon$ is
lesser than the $J/\Psi$, as the bottomonium states are smaller in size with larger masses than the charmonium states. 

\section{Effect of Auxilliary fields}
\label{sec:Effect of Auxilliary fields}
In this section, we extend the discussion of heavy quarkonium potential in presence of a localized action and it's medium effect on the potential  to check whether the localized action can contribute a linear term in the quarkonium potential. The restriction of the integration domain in the functional integral is realized by adding a non-local horizon term to the Faddeev-Popov action~\cite{Zwanziger:1988jt,Canfora:2015yia}.
The localization of the horizon non-locality was introduced in Refs.~\cite{Zwanziger:1989mf,Zwanziger:1988jt,Zwanziger:1992qr} by introducing Zwanziger ghost fields, $\{\phi_\mu^{ab},\bar\phi_\mu^{ab}\}$, a pair of commuting fields and $\{\om_\mu^{ab},\bar\om_\mu^{ab}\}$, a pair of anti-commuting fields. The Lagrangian is 
\begin{eqnarray*}
	L&=&L^{\text{YM}}-\bar c^a\partial ^\mu D_\mu^{ab} c^b+\frac{1}{2}\chi^{ab}_\mu \partial^\nu(D_\nu \chi^\mu)^{ab}\nn\\
	&+&\frac{i}{2}\chi^{ab}_\mu \partial^\nu(D_\nu \xi^\mu)^{ab}-\frac{i}{2}\xi^{ab}_\mu \partial^\nu(D_\nu \chi^\mu)^{ab}\nn\\
	&+&\frac{1}{2}\xi^{ab}_\mu \partial^\nu(D_\nu \xi^\mu)^{ab}
	- \bar \om^{ab}_\mu \partial ^{\nu}(D_\nu \om^\mu)^{ab}\nn\\
	&-&\frac{1}{\sqrt{2}} g f^{abc} \partial^\nu \bar \om_\mu^{ae}(D_\nu c)^b\chi^{ec\, \mu}-i \gm_G^2 f^{abc}A^{a\, \mu}\xi_\mu^{bc}\nn\\
	&-&\frac{i}{\sqrt{2}} g  f^{abc} \partial^\nu \bar \om_\mu^{ae}(D_\nu c)^b\xi^{ec\, \mu}-\frac{1}{2g^2}d N \gm^4,
\end{eqnarray*}
where $\phi_\mu^{ab}=\frac{1}{\sqrt{2}}(\chi_\mu^{ab}+i \xi_\mu^{ab})$.
The authors of Ref.~\cite{Gracey:2009mj} have calculated the static potential and discussed the nature of the potential with separation distance. The ghost fields are treated as internal lines in Feynman diagrams. Tree level and one-loop corrections are considered to calculate the static potential.
In this work, we have investigated the potential behavior in the presence of a thermal bath. The one loop potential in the Landau gauge is given by (see Ref.~\cite{Gracey:2009mj})
\begin{align}
&	\hspace{-.4cm}	{V}(p)=\nn\\
& -\frac{C_F g^2}{p^2}\Bigg\{ \bigg[ 1- \frac{C_A \gamma_G^4}{(p^2)^2}+ \mathcal{O}\bigg(\frac{\gamma_G^8}{(p^2)^4}\bigg)\bigg]\nn\\
&+ \bigg[ \bigg( \frac{31}{9}- \frac{11}{3}\log\left(\frac{p^2}{\Lambda^2}\right) \bigg) C_A\nn 
\end{align}
\begin{align}
&+ \frac{1}{2}\bigg(\frac{4}{3}\log\left(\frac{p^2}{\Lambda^2}\right)- \frac{20}{9}\bigg) N_f - \frac{2 \pi C_A^{3/2}\gamma_G^2}{p^2} \nn\\
&+ \bigg[ \bigg( \frac{79}{12}\log\left(\frac{p^2}{\Lambda^2}\right)+\frac{9}{8} \log\left(\frac{C_A \gamma_G^4}{(p^2)^2}\right)- \frac{1315}{72} \bigg) C_A^2 \nonumber\\
&+ \frac{1}{2}\bigg( \frac{40}{9}- \frac{8}{3}\log\left(\frac{p^2}{\Lambda^2}\right) \bigg)  N_f \bigg] \frac{C_A \gamma_G^4}{(p^2)^2}\nn\\
&+\mathcal{O}\bigg(\frac{\gamma_G^6}{(p^2)^3}\bigg) \bigg] \frac{g^2}{16} 
+ \mathcal{O}\left(g^4\right) \Bigg\}.
\end{align}
At non-zero temperature, we use an ansatz that the thermal-medium effect enters through the dielectric permittivity $\eps(p)$ such as~\cite{Agotiya:2008ie,Thakur:2020ifi}
$	\bar V(p)=V(p)/\eps(k).$

The dielectric permittivity, related to the static limit of $\Pi_L(p_0=0,p)=m_D^2$, is given by $\eps(p)=(1+m_D^2/p^2)$.
So, the in-medium quarkonium potential in real space becomes
	\begin{align}
\hspace{-.4cm}V(r) &=  \int \frac{d^3p}{(2\pi)^{3}} \,e^{i \bf p \cdot \bf r}\,p^2 \frac{V(p)}{p^2+m_D^2}\nn\\
=& - \frac{C_Fg^2}{4\pi} \left[4 \pi^2 + \frac{1}{36}(31 C_A -20 N_f T_f)g^2\right]\nn\\
&\left(\frac{1}{ r}+\frac{m_D^2r}{2}\right) -\frac{g^2 \alpha_s C_F}{72r} (11 C_A -4 N_f T_f)
\nonumber\\
&\times\bigg[6 [2 \gm_E+2\log (\Lambda r)] +3 r^2 m_D^2 [-3+ 2 \gm_E\nn\\
&+  2\log (\Lambda r) + 2 r^3 m_D^3  \log \left(\frac{m_D}{\Lambda}\right) \bigg] \nonumber \\
&+\frac{\pi r \alpha_s \gamma_G^2 }{12 m_D^2}\big[g^2 m_D^2 (-3C_A^{3/2}C_F + r m_D)\nn\\
&-8\pi C_A C_F r  \gamma_G^2\big]+ \frac{C_A C_F g^2 \alpha_s \gamma_G^4 r^2}{1728 m_D}\bigg[1920\nn\\
&-1315 C_A^2\left.+81 C_A^2 \log\left(\frac{C_A \gamma_G^4}{m_D^4}\right)\right]- \frac{C_A C_F g^2 \alpha_s \gamma_G^4 }{288 m_D^4 r}\nn\\
&\times \left[ (64-79 C_A^2)r^3 m_D^3 + 384 \sinh(m_D r)\right] \log\left(m_D^2/\Lambda^2\right), 
\label{eq:aux}
\end{align}
where $\gm_E= 0.57721566 ...$ is the Euler–Mascheroni constant. Equation~\eqref{eq:aux} is the medium-modified potential of Gribov plasma with auxiliary fields. It is apparent from Eq.~\eqref{eq:aux} that the linearly rising term ($\sim (+ve) r$) in the potential does not emerge with considering auxiliary fields at one loop order even after considering the medium effect.Instead of taking the given intuitive form of the dielectric permittivity, one can systematically calculate the thermal medium effect for these auxilliary fields, but the nature will remain same. One may also consider higher loop contributions to the HQ potential for improved results. 
\label{sec:summary}
\section{Discussion}
\label{sec:discussions}

The quarkonium potential and decay width are modified within Gribov scenario. It is clear from Eq.~\eqref{eq:vg} that the confining property is still absent in the quarkonium potential. It is evident from the fig.~\ref{fig_real} that at a fixed temperature, the heavy quarkonium static potential is marginally more screened than that obtained from pure HTL perturbation theory. 
In the temperature range we are interested in, the medium is non-perturbative and HTL result may not be appropriate to use. 
The Gribov parameter increases the debye mass and the potential becomes more screened as the debye radius is inversely proportional to the debye mass. 
	
\section{Summary and outlook}
In the present theoretical study, the real and imaginary parts of heavy quarkonium complex potential have been computed considering both the perturbative resummation using HTLpt and non-perturbative resummation using the Gribov-modified gluon and ghost propagators. 
We first obtained an effective gluon propagator in the presence of the Gribov parameter. The longitudinal and transverse part ($\Pi_L,\, \Pi_T$) of the effective propagator is then obtained from the one-loop gluon self-energy. The gluon self-energy gets the contribution from quark, gluon, and ghost loops. The dependence on the Gribov parameter comes from the gluon and ghost loops. Then we plotted the Debye mass. In our work, an asymptotically high-temperature form of the Gribov parameter $(\gm_G)$ has been considered. The real part of the potential is obtained from the Fourier transform of the static gluon propagator. The $Q\bar Q$ is more screened in the presence of the Gribov parameter. The imaginary part of the potential and the decay width are evaluated in Gribov plasma. The magnitude of the imaginary part of the potential increases with distance. Accordingly, the width increases with temperature. 
We have also discussed the medium effect on the HQ potential in the presence of auxiliary fields at one loop order. As also discussed in Ref.~\cite{Gracey:2009mj}, the linearly increasing potential in coordinate space is missing even with the additional auxiliary fields. The confining term in the quarkonium potential with Gribov action may appear beyond the leading order in the QCD weak-coupling expansion. 

\hspace{.6cm}
\section{Acknowledgment}  
N.H. is supported in part by the SERB-MATRICS under Grant No. MTR/2021/000939. The authors acknowledge Arghya Mukherjee for helpful discussions during the project.

\appendix
\section{Calculations of self-energy with Gribov parameter}
\label{app_selfenergy}
The expression of $J_1$ is
\bea
J_1&=&\int \frac{d^4 K}{(2 \pi)^4} \Delta(K) =\int \frac{d^4 K}{(2 \pi)^4} \frac{K^2}{K^4 + \gamma_G^4} \nn \\
&=&\frac{1}{4 \pi^2}\sum_{a=\pm} \int_0^{\infty} \frac{k^2 \, dk}{E_a(k)}n_B(E_a(k)),
\eea
where $E_a(k)= \sqrt{k^2 + i a \gamma_G^2}$.
The expression for $J_2$ can be written as
\bea
J_2(P)
&=& \frac{i \gamma_G^2}{4}\sum_{a,b=\pm 1} \int\frac{d^4 K}{(2 \pi)^4}\nn\\
&&\hspace{-1cm}\times\ \frac{a}{[\omega_n^2 + E_a^2 (k)][(\omega - \omega_n)^2 + E_b^2 (|\mathbf{p}-\mathbf{k}|)]}. 
\eea
The structure of $I_{\mu \nu}(P)$ looks like,
\bea
I_{\mu \nu}(P) &=& \int \frac{d^4K}{(2 \pi)^4} K_{\mu} K_{\nu} \Delta(K) \Delta(P-K)
\eea

To calculate the longitudinal $(\Pi_L)$ and transverse $(\Pi_T)$ part of the gluon self-energy, we need to calculate $I_{\mu\mu}$ and  $I_{44}$.
Now, 
\bea
I_{\mu\mu} &=&\int\frac{d^4K}{(2\pi)^4}K^2\Delta(K)\Delta(P-K)\\
I_{44} &=&J_1 - J_2 - J_3.
\eea 
Now, we calculate $J_3$ using HTL approximation. So, one can write
\bea
J_3 
&=&-\frac{1}{16}\int\frac{d^3k}{(2\pi)^3}\,k^2\sum_{s,s_1=\pm1}\sum_{a,b=\pm}\frac{ss_1}{E_a(k)E_b(|{\bf p-k}|)}\nn\\
&\times& \frac{1+n_B(sE_a(k))+n_B(s_1E_b(|{\bf p-k}|))}{i\omega-sE_a(k)-s_1E_b(|{\bf p-k}|)}
\eea
After analytic continuation and taking the static limit i.e. $i\omega\rightarrow p_0 = 0$ limit, we get,
\begin{align}
\text{Im}[J_3(p_0,\mathbf{p})] &=- i \pi \frac{p_0}{p} \frac{1}{8 T (2 \pi)^2} \int dk\, k^4 \nn\\ 
&\!\!\!\times \sum_{a,b=\pm 1} \frac{n_B(E_b(k))[1+n_B(E_b(k))]}{E_a(k)E_b(k)}.
\end{align}
Similarly for $J_2(p_0,\mathbf{p})$,
\begin{align}
\text{Im}[J_2(p_0,\mathbf{p})]&=- i \pi \frac{p_0}{p} \frac{i \gamma_G^2}{8 T (2 \pi)^2} \int dk \, k^2  \nn\\
&\!\!\!\times \sum_{a,b=\pm 1} \frac{n_B(E_b(k))[1+n_B(E_b(k))]}{E_a(k)E_b(k)}.
\end{align}

Now we compute the expression of $I_{00}^G(p_0,\mathbf{p})$ which appears in ghost loop.
\bea
I_{00}^G(p_0,\mathbf{p})&=&\int \frac{d^4K}{(2\pi)^4} \frac{k_0^2}{K^4 (P-K)^4} 
\eea
The real part contributes to the Debye mass. We also get the imaginary part from the Landau damping part,
\begin{eqnarray}
\lim_{p_0\to 0} \text{Im}[I_{00}^G(p_0,p)] &=& i \frac{\pi p_0}{p} \int_0^{\infty} \frac{dk}{(2\pi)^2} \frac{n_F(k)}{k^3}\nonumber\\
&=&\frac{\pi p_0}{p} \frac{7 \zeta(3)}{8 \pi^2 T^2}.
\end{eqnarray}
Hence, from Eq.~\eqref{pic} imaginary part of the self-energy from ghost loop looks like
\bea
\lim_{p_0\to 0} \text{Im}[\Pi_{00}^{(c)}(p_0,p)]=i \frac{\pi p_0}{p} C_A \frac{128^2 \gamma_G^4 }{32 N_c^2 g^2 T^2} 7 \zeta(3).\hspace{0.4cm}
\eea

	\bibliographystyle{elsarticle-num}
\bibliography{draft_HQ}

\end{document}